\documentclass[aps,prl,twocolumn,groupedaddress,showpacs]{revtex4}
\usepackage{graphicx}

\begin{document}


\title{Electron Temperature of Ultracold Plasmas}


\author{J.~L.~Roberts}
\altaffiliation[Present Address: ]{Physics Department, Colorado State University, Fort Collins, CO 80523}
\author{C.~D.~Fertig}
\author{M.~J.~Lim}
\altaffiliation[Present Address: ]{Department of Physics and Astronomy, Rowan University, Glassboro, NJ 08028}
\author{S.~L.~Rolston}
\altaffiliation[Present Address: ]{Department of Physics, University of Maryland, College Park, MD 20742}

\affiliation{Atomic Physics Division, National Institute of Standards and Technology, Gaithersburg, MD 20899}


\date{\today}

\begin{abstract}
We study the evolution of ultracold plasmas by measuring the electron temperature.  Shortly after plasma formation, competition between heating and cooling mechanisms drives the electron temperature to a value within a narrow range regardless of the initial energy imparted to the electrons.  In agreement with theory predictions, plasmas exhibit values of the Coulomb coupling parameter $\Gamma$ less than 1.
\end{abstract}


\maketitle

Knowledge of the electron temperature is critical to understanding the behavior of ultracold plasmas formed by the photoionization of laser-cooled atoms \cite{Killian1999}.  Until now, this important parameter had not been directly measured.  Simple consideration of the photoionization process (i.e. considering only single atoms in isolation) implies that the initial kinetic energy of the electrons in the plasma is proportional to the energy of the photoionization photon in excess of the ionization limit.  In this case, only the linewidth of the photoionization laser limits the lowest energy imparted to the electrons, and in principle plasmas with electron temperatures below 1$K$ could be created.  The relationship between the actual electron temperature and temperatures obtained from this simple consideration remains an open question.

Several strong heating and cooling processes in the plasma can have a radical effect on the temperature of the system.  Continuum lowering \cite{Mazevet2002,Hahn2001}, correlation-induced heating \cite{Kuzmin2002, Pohl2003}, three-body recombination to Rydberg atoms and the evolution (deexcitation) of the Rydberg atoms \cite{Killian2001,Robinson2000,Robicheaux2002,Robicheaux2003,Tkachev2002} are all predicted to result in heating of the electrons that increases with increasing density.  Conversely, the electrons possess thermal pressure that causes the plasma to expand \cite{Kulin2000}, and that in turn induces strong adiabatic cooling.  Evaporative cooling is certainly present as well.  All of these processes can act within the first few microseconds after the plasma is created, and so the evolution of the electron temperature will be determined by the balance of these (and perhaps other) heating and cooling processes.

The determination of the electron temperature will be useful in interpreting the results of future experiments with ultracold plasmas.  Measurements of the electron temperature test molecular dynamics simulations{'} \cite{Mazevet2002,Kuzmin2002} predictions of this fundamental plasma parameter.  Because electron-atom collisions such as Rydberg recombination and superelastic scattering influence temperature, atomic collision theory in plasmas can also be tested \cite{Hahn2001,Robicheaux2002,Robicheaux2003,Tkachev2002}.  Finally, a measurement of the electron temperature is necessary in order to determine to what extent, if any, the electrons of these ultracold plasmas are in the strongly coupled (i.e.~correlated) regime \cite{Ichimaru1982}.

The ratio of Coulomb energy to thermal energy, $\Gamma$, expresses the importance of correlations in the plasma.  This parameter is defined (for the electrons) as $\Gamma=\frac{1}{4\pi\epsilon_0}\frac{e^{2}/a}{k_{B}T}$ where $a=(\frac{3}{4\pi n})^{1/3}$ is the Wigner-Seitz radius, $e$ is the electron charge, $n$ is the average plasma density, and $T$ is the electron temperature.  For $\Gamma > 1$ (the strongly coupled regime), spatial correlations in the plasma develop and phase transitions are possible \cite{Ichimaru1982}.  It is predicted \cite{Hahn2001,Mazevet2002,Kuzmin2002,Pohl2003,Robicheaux2002,Robicheaux2003} that various heating effects in the plasma will each limit $\Gamma$ to be less than 1, so measuring $\Gamma$ also tests the validity of these theoretical treatments.  In general, strongly coupled neutral plasmas (distinct from strongly coupled non-neutral plasmas \cite{Bollinger1994,Hayashi1994,Thomas1994,Chu1994}) are rare and difficult to produce in the laboratory; ultracold plasmas could potentially be cold enough to reach the strongly coupled regime.

In this work we measure the electron temperature a short time after the plasma formation.  The creation of ultracold plasmas in our apparatus has been previously described in Ref. \cite{Killian1999}.  Briefly, we use a Magneto-Optic Trap (MOT) to collect and cool 4x$10^6$ metastable xenon atoms to a temperature of about $20 \mu$K.  The spatial density distribution is roughly gaussian with a typical rms radius $\sigma$ $\sim$ 250 $\mu$m.  We produce the plasma in a 10 ns two-photon excitation of up to 30\% of the initial sample, with the number $N$ of photoionized atoms and the initial electron energy $\Delta E$ controlled by the intensity and frequency of the photoionizing laser, respectively \cite{deltaE}.  We apply small electric fields ($\sim 0.3-4$ V/m) to the plasma during its evolution by controlling voltages on wire mesh grids ($>$95\% transmission) positioned above and below the plasma region.  Electrons that leave the plasma are directed to a microchannel plate (MCP) detector.  We determine the density and size of the plasma by measuring its radio-frequency (RF) response as in \cite{Kulin2000}, with a slight modification to the calibration used there.  In light of recent theoretical work \cite{Mazevet2002,Bergeson2003}, we now base our size calibration on the measured asymptotic expansion velocity of high $\Delta E$ ($\Delta E/k_B \geq 400K$), low $N$ (N $\leq 5$x10$^4$) plasmas \cite{RF}.

\begin{figure}
\includegraphics[width= 2.75 in]{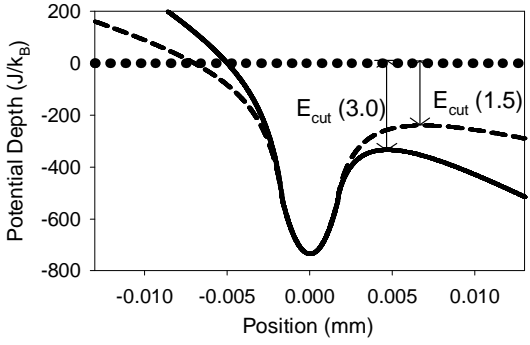}

\caption{\label{f1} Self-consistently calculated electron potentials shown as a function of position along the applied electric field direction, for $F$=1.5 V/m (solid line) and 3 V/m (dashed line) applied electric fields.  The relevant plasma parameters are $N_{ion}=10^6$, $N_{electron}=0.95N_{ion}$, $\sigma$ = 0.53mm, and $T$=40$K$.  The value of the parameter $E_{cut}(F)$ used in the calculation of the fraction of electrons spilled is determined from the saddle point as indicated.}
\end{figure}

After the photoionization pulse, some electrons rapidly escape as a result of the kinetic energy imparted during photoionization.  This results in an excess of ions as compared to electrons in the plasma region and a space charge develops.  This space charge in turn confines the remaining electrons in a potential well \cite{Killian1999}.  Elastic collisions rapidly redistribute the remaining electrons{'} energy into a thermal distribution \cite{Spitzer1962}.  Since the ions in the plasma are unconfined, the plasma expands outward in response to the thermal pressure of the electrons \cite{Kulin2000}.

\begin{figure}
\includegraphics[width= 2.75 in]{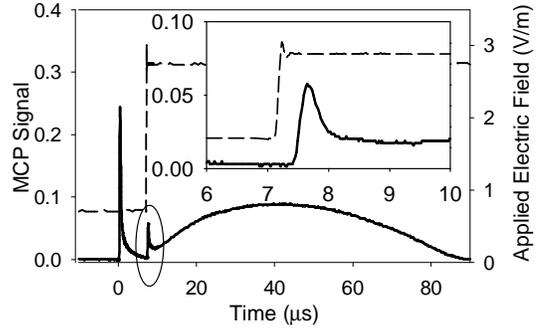}

\caption{\label{f2} MCP signal (solid line) showing a typical plasma evolution and spilling signal (region indicated in oval).  The inset shows the spilling part of the curve in more detail.  The magnitude of the applied electric field is also plotted (dashed line).}
\end{figure}

\begin{figure}
\includegraphics[width= 2.75 in]{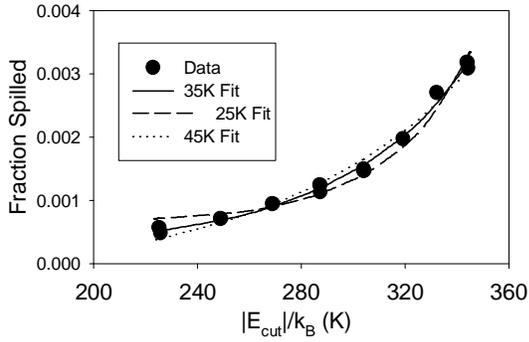}

\caption{\label{f3} Fraction of electrons spilled ($f$) vs. $E_{cut}$ for a typical set of data (filled circles).  Fits using the model based on Eq. (1) are shown for the temperatures that are indicated, where the 35$K$ curve is the best fit curve and the other two are shown for comparison.  Note that the value of $T$ is determined primarily from the curvature of $f$ vs.~$E_{cut}$.}
\end{figure}

Due to the low temperatures and small size of the plasma, the standard methods used to measure plasma temperatures are not applicable \cite{Hutchinson1990}.  We developed a method in which we probe the energy distribution of the electrons to measure $T$.  After a desired amount of evolution time, an electric field is turned on.  This field suddenly lowers the lip of the potential well confining the electrons, and energetic electrons become unconfined (see Fig. \ref{f1}).  These electrons then spill from the plasma and strike the MCP.  Figs. \ref{f2} and \ref{f3} show a typical spill signal and a plot of the fraction of electrons spilled as a function of the electric field amplitude $F$, respectively.  In order to avoid having to dynamically model the changing confinement during the spill, the fraction spilled $f$ is kept small enough so that the ratio $N_{electron}/N_{ion}$ is not altered substantially.

A model is necessary in order to extract the temperature from the measured spilling curves.  We developed a method to solve for the electron density distribution given the ion density, electron number, and temperature.  Since the electrons{'} confining potential is influenced by the electrons themselves, their density distribution is calculated self-consistently, starting from an approximate distribution and iteratively improving that distribution until it converges to a self-consistent result.  While this self-consistent calculation allowed us to calculate $f$ as a function of $F$, we found that $f$ calculated this way was highly sensitive to the details of the ion density distribution.  We did not want the measured temperature to be contingent on a simulation of the motion of the ions in the plasma (especially since the density of atoms in our MOT is not perfectly spherically symmetric or gaussian).  Therefore, we switched to a more robust method that does not depend explicitly on the ion density distribution.  This method relies on the fact that at positions far enough away from the center of the plasma, the electron potential is $U(r)=-\frac{(N_{ions}-N_{electrons})e^2}{4 \pi \epsilon_0} \frac{1}{r}$, regardless of the ion and electron density distributions.  The fraction spilled as a function of $F$ is then modeled using the following expression: 
\begin{equation}
f=\beta+\lambda\int^{E_{cut}(F)}_{E_{cut}(F_0)} D(\epsilon) \exp(-\frac{\epsilon}{k_BT}) d\epsilon
\end{equation}
where $D(\epsilon)$ is the density of states calculated for the plasma's electron potential in its asymptotic limit, $\lambda$ and $\beta$ are fit parameters corresponding to the chemical potential and a background subtraction, $E_{cut}(F)$ is determined by the magnitude of $F$ as shown in Fig. \ref{f1}, and $E_{cut}(F_0)$ is determined in the same way for $F_0$, which is the bias electric field present before the spilling field is turned on.  As long as the saddle point associated with $E_{cut}$ is far enough from the center of the plasma, Eq. (1) will be a good approximation of the number of electrons spilled in response to lowering the potential barrier.  

We tested the validity of this model in several ways.  First, we compared values of $T$ derived from fits to \mbox{Eq. (1)} with $T$ determined from the more sophisticated self-consistent model for a variety of different ion density distributions.  The values of $T$ from the two different methods matched to better than 20\% for the plasma parameters examined in this work.  The form of $D(\epsilon)$ was varied and found to have little effect on the fit temperature \cite{exponential}.  Finally, experiments were performed where RF pulses were applied to impart a variable amount of heat.  The temperatures determined using Eq.~(1) shortly after these pulses scaled linearly with applied RF power \cite{scaling}.

In order to accurately calculate $f$ in both the self-consistent model and in the model represented by Eq. (1), we must include the screening of the external field by the plasma.  This screening is approximated by the use of a ``screening radius,'' within which the external electric field does not penetrate.  The resulting induced dipole field is then included in the calculation of $E_{cut}$.  We estimate values for the screening radius by calculating the first-order correction, due to the presence of the electric field, to the electron density distribution self-consistently determined in the absence of the field. 

Temperature measurements were taken over a range of ion numbers (0.1-2x10$^6$), at several different values of $\Delta E$, and at different plasma evolution times.  The plasma evolution times that can be studied are constrained to be from $\sim$3-8 $\mu s$ for higher $\Delta E$ and from $\sim$4-15 $\mu s$ for lower $\Delta E$.  At early times, the electrons in the prompt peak distort the spilling signal.  At later times, the plasma expands to sizes large enough so that the saddle points induced by the spilling field are inside the $\frac{1}{r}$ part of the potential, violating a central assumption of our model and thus rendering it inapplicable.

\begin{figure}
\includegraphics[width= 2.75 in]{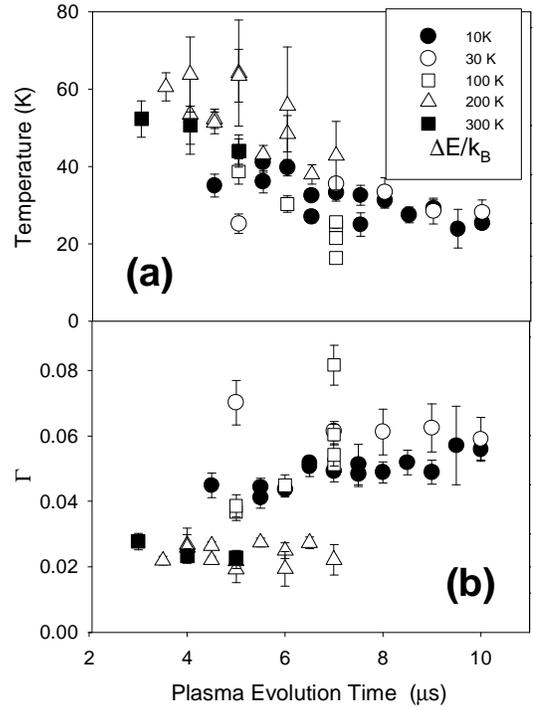}
\caption{\label{f4} (a) Measured temperatures vs.~plasma evolution time for different values of $\Delta E/k_B$ as indicated in the legend.  These data have $N_{ion} \sim 10^6$, 0.4mm $\leq \sigma \leq$ 0.6mm for $\Delta E/k_B \leq 100K$ and 0.4mm $\leq \sigma \leq$ 0.7mm for $\Delta E/k_B > 100K$.  This size range was chosen because it corresponds to the plasma evolution time just after the initial electrons have escaped from the plasma.  Times 0.5-2 $\mu$s earlier (depending on $\Delta_E$) and a few microseconds later were also measured.  The error bars show only the statistical uncertainty.  (b) Corresponding $\Gamma$ for the data shown in Fig. 4a.  The $\Delta E/k_B = 100K$ data do not continue to increase but rather level out at later evolution times.}
\end{figure}

Figure \ref{f4} shows the measured values of $T$ (Fig. \ref{f4}(a)) and $\Gamma$ (Fig. \ref{f4}(b)) vs.~plasma evolution time for various values of $\Delta E$.  Two features in Fig. \ref{f4} are immediately apparent.  First, cooling is observed as the plasma expands, as is expected from the dynamics of the expansion \cite{Kulin2000}.  Second, the temperatures observed for the lower values of $\Delta E$ are greater than $\Delta E$ itself, clearly indicating the importance of heating in the early stages of the plasma.  Indeed, the range of temperatures observed falls in a remarkably narrow band given the factor of 30 range in the value of $\Delta E$.  As the number of ions is varied, the same qualitative features are present.  The values of $T$ decrease as $N$ is reduced, but in such a way that $\Gamma$ does not increase substantially.  For the most favorable conditions (late times, $\Delta E/k_B =$ 30$K$-100$K$, low $N$) values of $\Gamma$ in the range of 0.1-0.15 were consistently achieved, and thus these plasmas are not in the strongly coupled regime at these times.  It is still possible that $\Gamma$ increases above 1 at much later times where we cannot apply our measurement methods.  However, we do not see evidence for rapidly increasing $\Gamma$ vs.~time at the later evolution times we can observe.

Several systematic uncertainties have to be taken into account.  The dominant uncertainties are in the calibration of the ion number (35\%), in the appropriate value of the screening radius, and in the calibration of the plasma density.  When combined in quadrature, these uncertainties produce a 70\% overall systematic uncertainty in the measurement of $T$.

Several assumptions are also made in the analysis:  that the highest-energy electrons are in thermal equilibrium with the rest of the electrons; that $f$ is proportional to the number of electrons with energy greater than $E_{cut}$; and that the implicit truncation in the thermal distribution due to the finite potential depth does not affect its Maxwell-Boltzmann character at the energies corresponding to the values of $E_{cut}$ used.  By examining spilling curves and observing the flux of escaping electrons after the spilling peak, we determined that the plasma evolution times at which we took data were four times or more longer than the characteristic time that it took for elastic collisions to fill the spilled energy levels, supporting the first assumption.  The last two assumptions are consistent with our observations based on the comparison of spilling data with different ranges of $E_{cut}$ (i.~e.~cutting more or less deeply for the same experimental parameters), in which we observed no significant shift of $T$ as a function of the range of $E_{cut}$.

The observation that $\Gamma$ is limited to values less than 1 is consistent with theory predictions \cite{Hahn2001,Mazevet2002,Kuzmin2002,Pohl2003,Robicheaux2002,Robicheaux2003}.  Given the previously measured importance of Rydberg atom formation and evolution \cite{Killian2001} and the range of plasma evolution times studied in this work, comparing our results to the $\Gamma \sim 0.2$ prediction of Refs. \cite{Robicheaux2002,Robicheaux2003} is more direct than comparison to the other predictions cited.  For the relatively early plasma evolution times studied in this work, we measured values of $\Gamma$ that are significantly less than predicted by Refs. \cite{Robicheaux2002,Robicheaux2003}, and indeed are less than what would be naively implied by the other predictions \cite{Hahn2001,Mazevet2002,Kuzmin2002,Pohl2003} as well.  Measurement of the full time evolution of and $T$ and $\Gamma$ remains a challenge that will require a new measurement technique.  In the future, it is possible that measurements of the Rydberg distribution as a function of plasma evolution time can be used to measure temperatures over a larger fraction of the plasma lifetime.

In conclusion, we have developed a method of measuring the temperature of ultracold plasmas shortly after the plasmas are created.  This is done by measuring the response of the plasma to electric fields and thereby obtaining the energy distribution of the highest-energy electrons.  Significant cooling was observed when the initial kinetic energy of the electrons was large, and sig\-nif\-i\-cant heating was observed when low initial kinetic energies were imparted, driving the temperatures into a relatively narrow range.  The electron components of the ultracold plasma were never observed to be in the strongly coupled regime, confirming predictions by theory.  The colder ions may yet be in the strongly coupled regime, a possibility currently being studied by another group \cite{Killian2003}.

\begin{acknowledgments}
The authors acknowledge useful discussions with Francis Robicheaux during the course of this work.  Robert Fletcher provided assistance in the collecting and analyzing the data.  Scott Bergeson, Tom Killian, and Simone Kulin also contributed to the early stages of these experiments.  This work was funded in part by the ONR.
\end{acknowledgments}



\begin{thebibliography}{16}
\bibitem{deltaE}
The value of $\Delta E$ is determined from the photoionizing laser frequency such that $\Delta E=\hbar
\omega_{laser}-\hbar \omega_{IP}$, where $\omega_{laser}$ corresponds to the laser frequency and 
$\omega_{IP}$ is the frequency of the laser that will excite the electron exactly to the ionization
threshold.
\bibitem{RF}
The low $N$ and high $\Delta E$ condition is selected to minimize the amount of
density-dependent heating with respect to the initial kinetic energy imparted to the
electrons.  The expansion rate is then determined from $\Delta E$
 assuming energy
conservation.
\bibitem{exponential}
The insensitivity to the form of $D(\epsilon)$ is not surprising as the exponential Boltzmann
factor changes much more rapidly with energy than $D(\epsilon)$.  $D(\epsilon)$ for a $\frac{-1}{r}$
potential scales as $\epsilon^{-5/2}$.
\bibitem{scaling}
In the absence of a concrete prediction for the scaling of plasma temperature with applied
RF amplitude, the linear scaling of temperature with pulse amplitude is reasonable, but not
explicitly predicted.
\end{thebibliography}
\end{document}